\documentclass[reprint,aps,pre,showpacs,amsmath]{revtex4-1}
\usepackage{amssymb}
\usepackage{amsmath}
\usepackage{graphicx}

\begin{document}


\title{Locking of periodic patterns in Cahn-Hilliard models for Langmuir-Blodgett transfer} 



\author{Markus Wilczek}
\email[]{markuswilczek@uni-muenster.de}
\author{Svetlana V. Gurevich}
\affiliation{Institute for Theoretical Physics, University of M\"unster, Wilhelm-Klemm-Str. 9, 48149 M\"unster, Germany}


\date{\today}

\begin{abstract}
The influence of a periodic spatial forcing on the pattern formation in a generalized Cahn-Hilliard model describing Langmuir-Blodgett transfer is studied. The occurring synchronization effects enable a control mechanism for the pattern formation process. In the one-dimensional case the parameter range in which patterns are created is increased and the patterns' properties can be adjusted in a broader range. In two dimensions, one-dimensional stripe patterns can be destabilized, giving rise to a multitude of novel complex two-dimensional structures, including oblique and lattice patterns.
\end{abstract}


\maketitle 

\section{Introduction}
Locking and synchronization phenomena are ubiquitous in diverse areas, starting from physiological systems like oscillators in the human body \cite{glass2001synchronization,pikovsky2003synchronization} to technical applications such as chaotic pulsed lasers, magnetic nano-oscillators, or modern power grids \cite{sugawara1994observation,kaka2005mutual,filatrella2008analysis}. A rather beneficial utilization of synchronization effects is the control of pattern formation processes. While such systems exhibit natural spatial and temporal frequencies, they can be entrained to an external forcing, providing an additional control mechanism. Although the term synchronization is more commonly known in the context of temporal oscillations, there are numerous systems where spatial patterns exhibit synchronization or wavenumber locking to an external spatial forcing, e.g., in spatially forced chemical systems \cite{dolnik2011locking} or convection \cite{lowe1983commensurate,mccoy2008self,freund2011rayleigh}. Extensive theoretical studies have also been made in the context of Turing patterns \cite{rudiger2003dynamics}, Ginzburg-Landau type of equations \cite{coullet1986commensurate}, Swift-Hohenberg equations \cite{manor2008wave} and phase separation phenomena \cite{Krekhov2004Phase,Krekhov2009Formation}. Among others, the control via synchronization can be utilized in coating processes with thin layers \cite{krebs2009fabrication, li2012structure}. The coverage of substrates with thin layers of organic molecules finds various applications today, for example in the creation of sensor devices \cite{swalen1987molecular} or organic 
transistors and light emitting diodes \cite{dimitrakopoulos2001organic}. The employment of self-organization phenomena for the creation of such thin layers facilitates the production of structured layers \cite{bolognesi2005micro,chen2007langmuir}. The effective use of self-organization phenomena of course necessitates extensive control over the whole process. One of the common methods to control pattern formation processes is the use of prestructures that enable locking effects to occur \cite{gau1999liquid,qin1999fabrication,chen2000selective,checco2006liquid,koepf2011controlled}. If these effects are robust, the requirements regarding the accuracy of the rest of the process are lower and therefore in practice more easily achievable.\par
\begin{figure}[ht]
  \centering
 \includegraphics[width=0.5\textwidth]{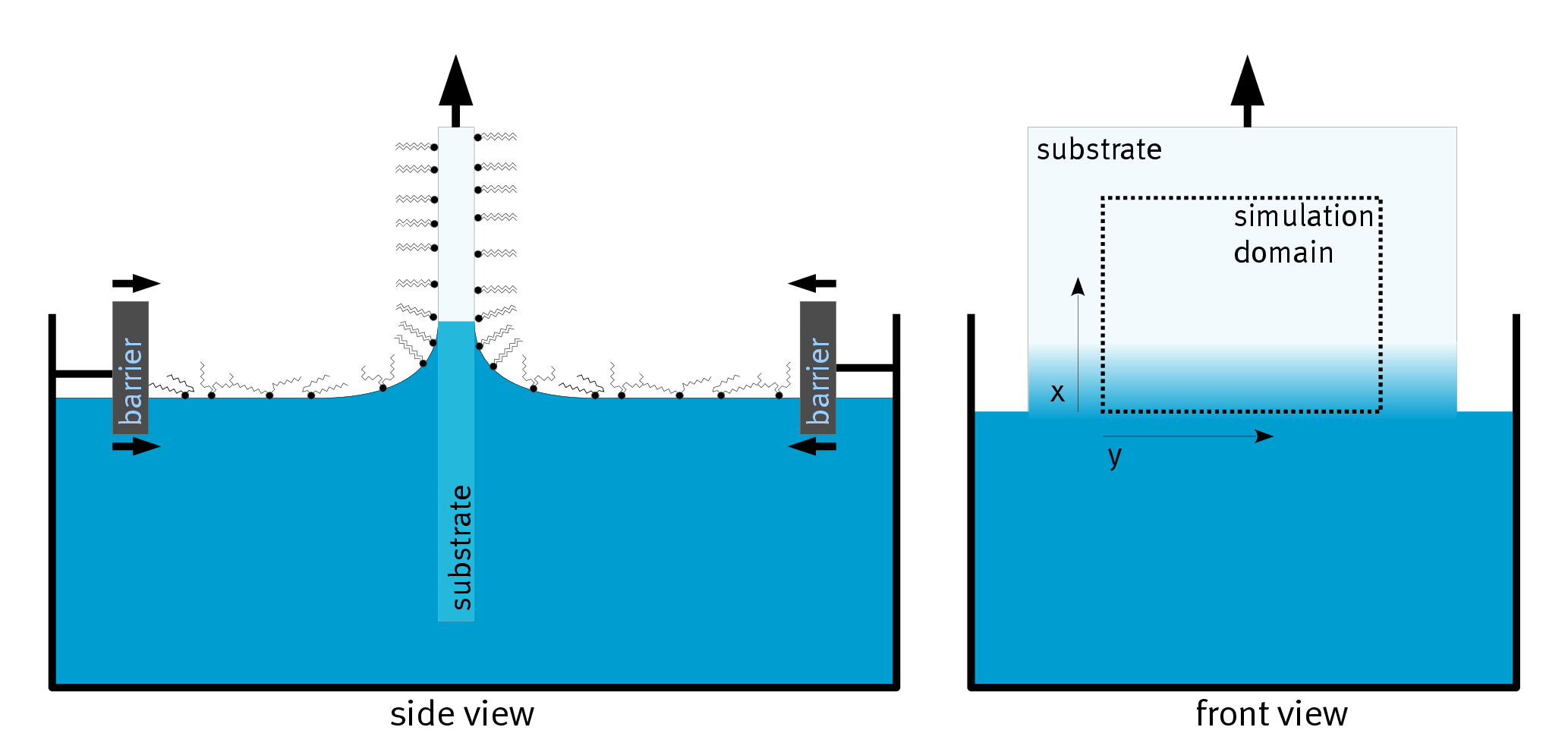}
 \caption{Left panel: Sketch of the experimental setup for Langmuir-Blodgett transfer. A substrate is pulled out of a trough filled with water, on which a floating monolayer of amphiphilic molecules is prepared. Movable barriers compress the monolayer during the transfer to keep the area density constant. Right panel: The front view illustrates the frame of reference and the coordinate system used in the numerical simulations.}
 \label{fig:LBtrough}
\end{figure}
Here we are interested in a description of synchronization effects in a generalized Cahn-Hilliard model \cite{koepf2012substrate} which was introduced to describe self-organized patterns arising by Langmuir-Blodgett (LB) transfer \cite{chen2007langmuir,blodgett1935films} onto prestructured substrates. The experimental setup for LB transfer consists of a trough filled with water, on which a floating monolayer of amphiphilic molecules, such as DPPC (dipalmitoylphosphatidylcholine), is prepared (see Figure \ref{fig:LBtrough}, left panel). A substrate is then pulled out of the trough, leading to a transfer of the floating monolayer onto the substrate. Movable barriers on the surface of the water are used to keep the area density of the monolayer constant, even if molecules are carried away on the substrate. As the monolayer is confined to the surface of the water, it constitutes a truly two-dimensional gas or liquid. The phase of the monolayer depends on its density and can therefore be controlled by the movable 
barriers. In the experiments that we are interested in \cite{chen2007langmuir,Gleiche2000Nanoscopic,spratte1994physisorption}, the monolayer is in a low density liquid-expanded (LE) phase. However, during the transfer, the monolayer is subject to a short-range interaction with the substrate, the so-called substrate-mediated condensation (SMC) \cite{spratte1994physisorption,riegler1992structural,graf1998molecular,spratte1994steady}. This effect lowers the coexistence pressure of the LE phase with the more dense liquid-condensed (LC) phase in the vicinity of the substrate. Therefore a phase transition of the monolayer into the LC phase is energetically favored as soon as it is transferred onto the substrate. This is experimentally 
observed if the substrate is pulled out of the trough sufficiently slow. For higher transfer velocities, however, the condensation of the monolayer does not occur uniformly but periodically, leading to the transfer of patterns consisting of domains in the LC phase alternating with domains in the LE phase. The patterns that can be obtained are highly regular stripe and lattice patterns \cite{chen2007langmuir}. As a main control parameter, the transfer velocity determines the type of the pattern and its properties like wavelength and orientation.\par
A way to gain more control over the pattern formation process is the use of prestructured substrate. In the following, we will consider substrates that have a periodic prestructure, which means certain properties of the substrate vary with a well defined spatial frequency. This introduces a periodic forcing, which can change the properties of the generated patterns, depending on, e.g., the strength of the forcing. This in turn is related to the contrast of the prepattern, i.e. how strong a certain property of the substrate varies along the prestructure. If the contrast is strong enough, the pattern formation process synchronizes to the prestructure, resulting in a perfect control over the produced structures.\par
This paper is organized as follows: Section 2 contains a derivation of the generalized Cahn-Hilliard model. In Section 3 the results of this model in the case of a transfer onto homogeneous substrates are briefly discussed. The results for the transfer onto prestructured substrates are presented in Section 4 for the one-dimensional case and in Section 5 for the two-dimensional case, respectively.
\section{Model equations}
Since the monolayer is in a stable LE phase before the transfer, the formation of domains in the LC phase can only occur in the vicinity of the substrate. Therefore the relevant part of the experiment that needs to be described by a theoretical model is confined to the meniscus, where the water layer between the monolayer and the substrate becomes thin. Under this assumption, the transfer process can be well described by the dynamics of the water layer in a lubrication approximation \cite{oron1997long} coupled to the dynamics of the floating monolayer on its surface \cite{craster2009dynamics, matar1997linear}. Such a model has been developed in \cite{koepf2010pattern} and proven to be able to describe most phenomena occurring during Langmuir-Blodgett transfer onto homogeneous substrates, as well as onto prestructured substrates \cite{koepf2011controlled}. However, the results of this model indicate that the dynamics of the water layer only has a minor impact on the pattern formation process, which is 
dominated by the dynamics of the floating monolayer undergoing a phase decomposition. Therefore a reduced model can be derived, in which the water layer is assumed to be static and its shape only enters parametrically \cite{koepf2012substrate}. The evolution equation for the density of the monolayer then has the form of a generalized Cahn-Hilliard equation. While in \cite{koepf2012substrate} such a model has been presented on the basis of the full thin film model, here we introduce the model starting from the general Cahn-Hilliard equation \cite{novick1984nonlinear} and introducing the contributions specific to the case of LB transfer. It has to be emphasized that, while this minimal theoretical model for LB transfer has proven to be able to capture the main features of pattern formation occurring in the experiments, we do not intend to make quantitative predictions. Therefore we do not give a specific scaling of the dimensionless quantities used in the equation and also do not derive parameters from 
experimental data. In contrast, we want to facilitate the comparison of the results with results obtained in similarly general models, like in the Swift-Hohenberg equation \cite{manor2008wave} or in reaction-diffusion systems \cite{dolnik2011locking}.\par 
The Cahn-Hilliard equation for the concentration $c(\mathbf{x},t)$ of the monolayer in one ($\mathbf{x}=x \in \Omega_1 \subset \mathbb{R}$) or two ($\mathbf{x}=(x,y) \in \Omega_2 \subset \mathbb{R}^2$) dimensions reads
\begin{equation}
 \frac{\partial }{\partial t} c(\mathbf{x},t) = \nabla \cdot  \left( M \nabla \frac{\delta F(c)}{\delta c} \right),
 \label{eq:CH}
\end{equation}
with the mobility $M$ and the free energy $F(c)$ given by
\begin{equation}
 F(c) = \int \frac{1}{2} \left( \nabla c \right)^2 -\frac{1}{2} c^2 + \frac{1}{4} c^4 + \mu \zeta (\mathbf{x}) c \ d\mathbf{x}.
 \label{eq:freeenergy}
\end{equation}
The free energy (\ref{eq:freeenergy}) includes the Cahn-Hilliard contribution due to spatial inhomogeneities \cite{cahn1958free}, and a double-well approximation for the free energy of the uniform system, which is justified in the vicinity of the LE-LC phase transition of the monolayer \cite{koepf2010pattern}. Here, $\mu$ is a coefficient regulating the strength of the SMC, which is spatially varying with the function $\zeta(\mathbf{x})$. For $\mu=0$ the two minima of the free energy $c=\pm 1$ have equal depth and correspond to a monolayer in the pure LE phase ($c=-1$) or in the pure LC phase ($c=+1$). For $\mu>0$ the double well of the free energy has a skewness favoring the LC phase. The form of $\zeta(\mathbf{x})$ reflects the characteristics of the liquid layer between the monolayer and the substrate located at the meniscus. Here we use the form
\begin{equation}
 \zeta(\mathbf{x}) = -\frac{1}{2}\left( 1+\tanh \left( \frac{x-x_\mathrm{s}}{l_\mathrm{s}}\right) \right),
\end{equation}
which ensures a smooth transition from no influence of the SMC ($\zeta(\mathbf{x})=0$) before the meniscus ($x<x_\mathrm{s}$) to a fixed value ($\zeta(\mathbf{x})=-1$) after the meniscus ($x>x_\mathrm{s}$), which is located at $x_\mathrm{s}$. The width of the transition region is determined by $l_\mathrm{s}$. While we will use the hyperbolic tangent shape in the following, the concrete shape of $\zeta(\mathbf{x})$ only has a minor impact on the simulations.\par
The transfer process is included into the model through an additive advective term $\mathbf{v} \cdot \nabla c$ with the transfer velocity $\mathbf{v} = (v,0)$. Incorporating these contributions and assuming the mobility to be constant $(M=1)$, the final model reads
\begin{equation}
 \frac{\partial}{\partial t} c(\mathbf{x},t) =   \nabla \cdot \left[  \nabla \left( -\Delta c - c + c^3 + \mu \zeta(\mathbf{x}) \right) - \mathbf{v} c\right].
\label{eq:model}
\end{equation}
This equation is solved numerically on a one- or two-dimensional domain ($\Omega_1 = [0,L]$ and $\Omega_2=[0,L]\times [0,L]$, respectively) with the boundary conditions
\begin{equation}
 \left. c \right|_{x=0} = c_0, \quad   \left. \frac{\partial}{\partial x} c \right|_{x=L} = 0,    \quad  \left. \frac{\partial^2}{\partial x^2} c \right|_{x=0,L} = 0,
\label{eq:boundary_conditionsx}
\end{equation}
\begin{equation}
 \left. c \right|_{y=0} = \left. c \right|_{y=L}.
\label{eq:boundary_conditionsy}
\end{equation}
The  boundary at $x=0$ reflects the role of the Langmuir-Blodgett trough supplying a constant density $c_0$ of the monolayer on the surface of the water and therefore in front of the meniscus. Ideally, the boundary at $x=L$ should not influence the outflow of the concentration $c$ with $\mathbf{v}$, which is hard to realize. However, the conditions given in (\ref{eq:boundary_conditionsx}) are nearly non-reflective, restricting the region of their impact close to the boundary. On a sufficiently large simulation domain, the influence onto the region of interest can therefore be safely neglected. The boundary conditions in $y$ direction (\ref{eq:boundary_conditionsy}) are chosen to be periodic in two-dimensional simulations (see Fig. \ref{fig:LBtrough} (right panel) for a clarification of the coordinate system). The simulations were performed using an adaptive Runge-Kutta 4(5) time-integration scheme and finite difference evaluations of the spatial derivatives on a grid of 500 ($384 \times 384$ in 2D) points. 
The methods were implemented using the NVIDIA CUDA framework \cite{CUDAGuide} for computations on graphics processors.\par

\section{Transfer onto homogeneous substrates in 1D}
\begin{figure}[ht]
  \centering
 \includegraphics[width=0.5\textwidth]{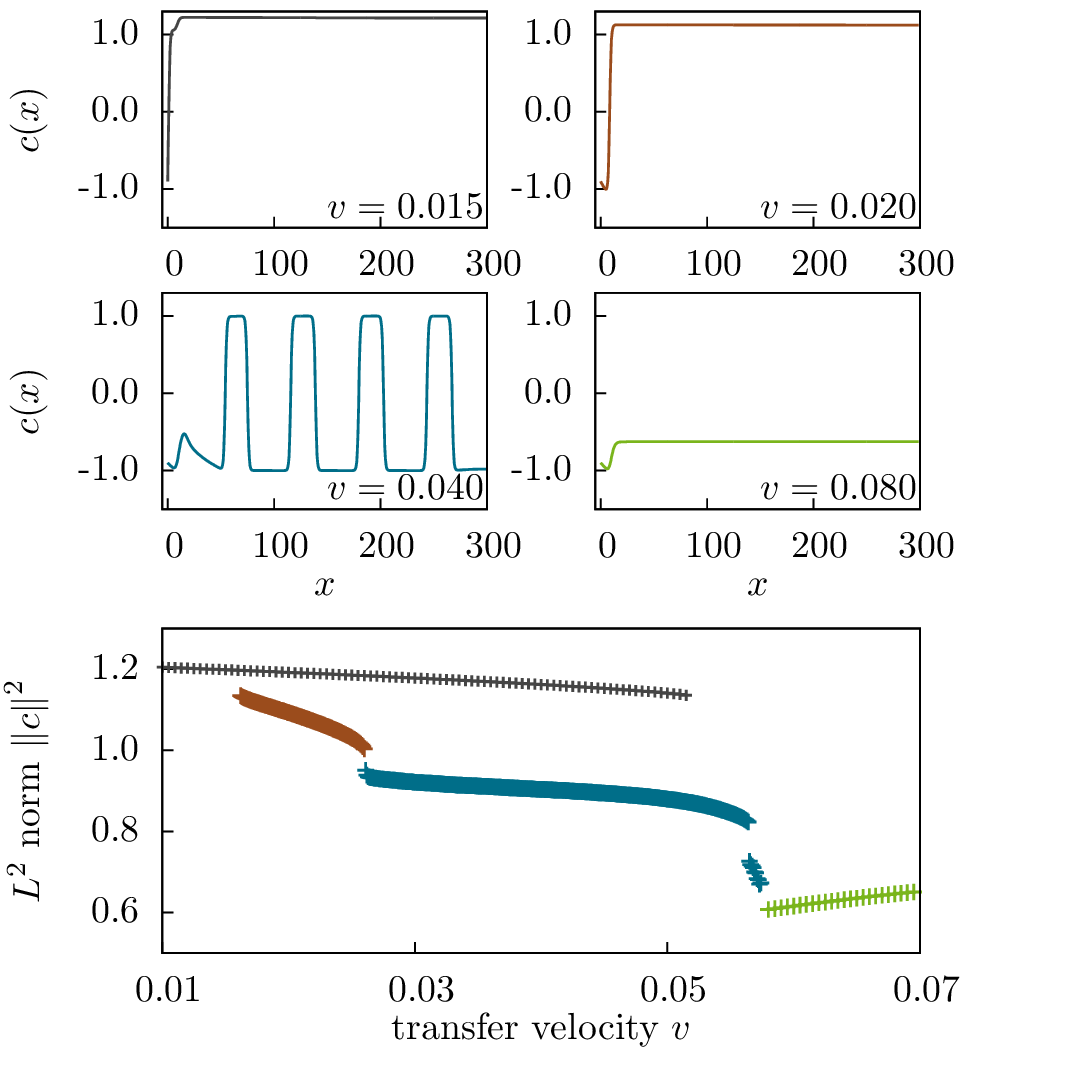}
 \caption{Overview of stable stationary and periodic solutions of (\ref{eq:model})-(\ref{eq:boundary_conditionsx}) in one dimension (bottom panel). The solution types are shown as branches in a $L^2 = \left\Vert c(x)\right\Vert^2 := \sqrt{\frac{1}{L}\int_0^{L} c(x)^2 \ \mathrm{d}x}$ versus $v$ diagram. For each branch an exemplary solution is shown (four top panels).}
 \label{fig:branches}
\end{figure}

The transfer onto a homogeneous substrate in one dimension has already been discussed in details in \cite{koepf2012substrate}. There, four stable solution types of (\ref{eq:model})-(\ref{eq:boundary_conditionsx}) have been identified: Two solution types corresponding to the transfer of a homogeneous LE layer for low velocities, one solution type corresponding to a homogeneous LC layer for high velocities, and one corresponding to stripes of alternating LE and LC domains parallel to the meniscus for intermediate velocities. These results, which can be obtained by direct numerical simulations, are shown in Figure \ref{fig:branches}. Each solution type corresponds to a branch in a diagram where the $L^2$ norm of the solutions is plotted against the transfer velocity. In \cite{koepf2012substrate}, such a diagram was augmented by unstable stationary solutions which were traced using continuation methods, resulting in a complete bifurcation diagram for the one-dimensional system. In this diagram, the unstable 
stationary solutions exhibit a heteroclinic snaking behavior connecting the stable stationary solutions corresponding to homogeneous LE and LC layer transfers. One of the unstable solution branches is also connected to the stable LE branch by the branch corresponding to the periodic solutions (Figure \ref{fig:branches}, blue line) generating stripes parallel to the meniscus. This branch emerges in a homoclinic bifurcation at low velocities, while at high velocities, it ceases in a sequence of subcritical Hopf bifurcations.\par
\begin{figure}[ht]
  \centering
 \includegraphics[width=0.5\textwidth]{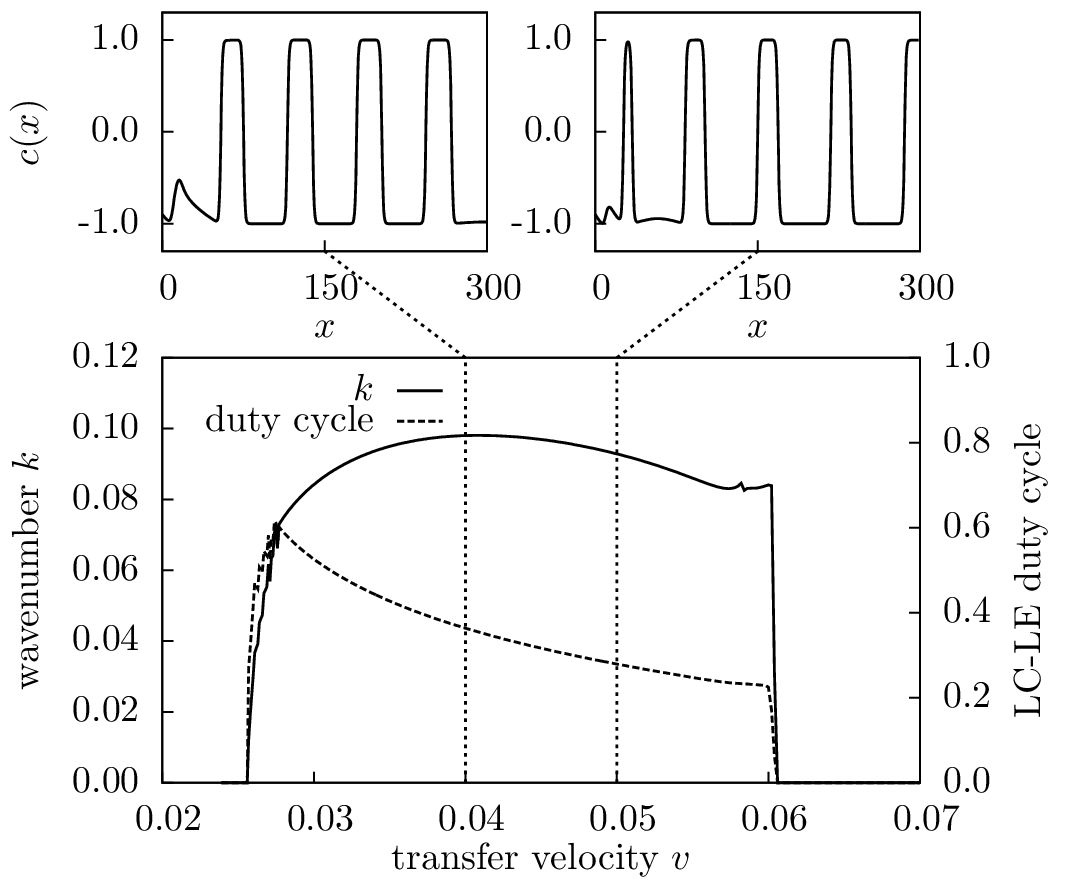}
 \caption{Wavenumber $k$ (solid line) and duty cycle (dashed line) of generated patterns during a transfer onto a homogeneous substrate in one dimension in dependence on the transfer velocity $v$ for $c_0=-0.9$ (bottom). Two snapshots of exemplary solutions for $v=0.04$ and $v=0.05$ are shown (top).}
 \label{fig:k-vs-v}
\end{figure}

To be able to discuss the influence of prestructured substrates on the dynamics of periodic solutions, we will start with a discussion of the properties of the one-dimensional case of a transfer onto a homogeneous substrate, where the generated pattern can be characterized by the wavenumber $k$ and the duty cycle, i.e. the ratio between the width of the stripes in the LC phase and the wavelength. The dependence of the wavenumber $k$ and the duty cycle on the transfer velocity $v$ is shown in Figure \ref{fig:k-vs-v}. Up to a certain threshold velocity, a homogeneous layer in the LE phase is transferred. Above this threshold, domains in the LC phase alternating with domains in the LE phase arise at the location of the meniscus and are then carried away with the transfer velocity $v$. That is, temporal oscillations of the concentration at the meniscus translate to the resulting spatial patterns on the substrate. At the onset of pattern formation, the wavenumber $k$ steeply increases with $v$ up to a 
maximum for intermediate transfer velocities after which the wavenumber decreases again. Therefore patterns with the same wavenumber are created for different velocities. The patterns are not identical, however, because the duty cycle of the generated pattern is monotonically decreasing with increasing speed. Above a certain threshold velocity, a homogeneous layer in the LC phase is transferred, corresponding to $k=0$. We define the velocity interval between these threshold velocities as the \emph{patterning regime}. At the upper limit of the patterning regime, the location where new stripes are formed is carried further and further into the domain, and the patterning regime therefore ends as soon as this location is outside the simulation domain. The upper threshold velocity is therefore dependent on the actual domain size. \par
\begin{figure}[ht]
  \centering
 \includegraphics[width=0.5\textwidth]{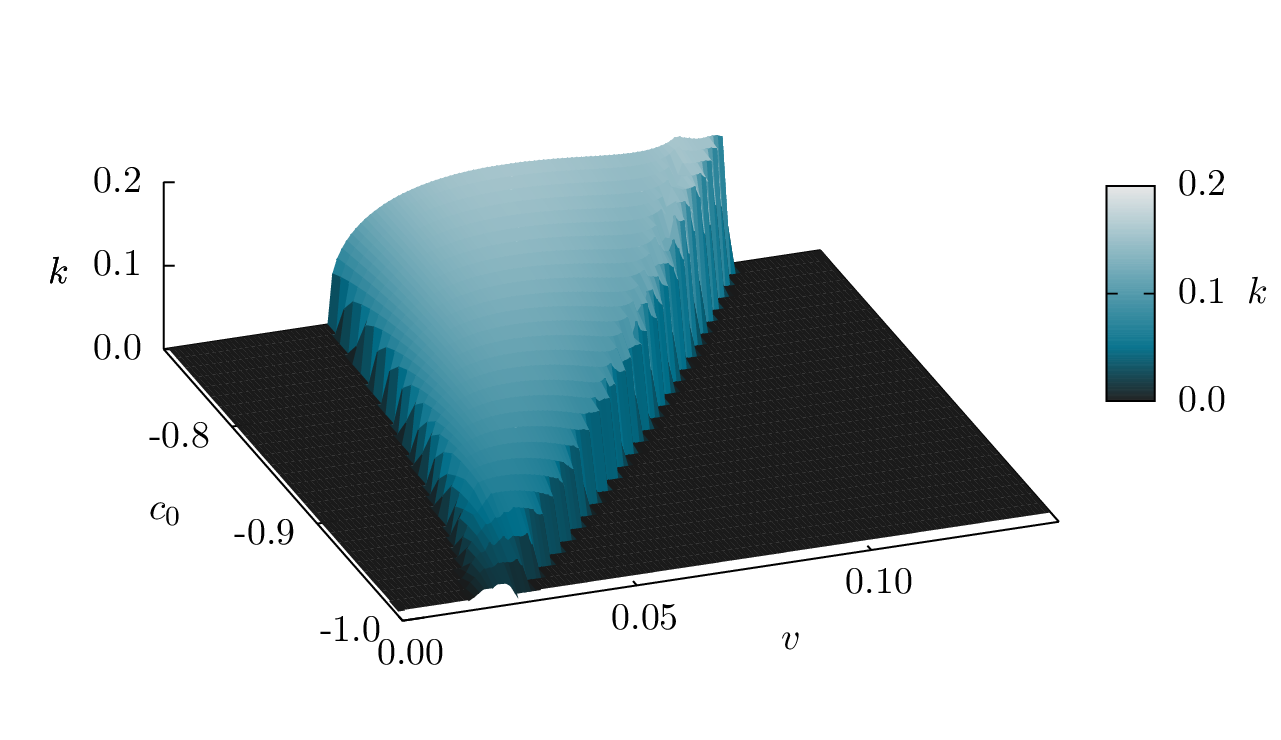}
 \caption{Wavenumber $k$  of generated patterns during a transfer onto a homogeneous substrate in one dimension in dependence on the transfer velocity $v$ and the boundary concentration value $c_0$. The patterning regime broadens and shifts towards higher velocities for increasing $c_0$.}
 \label{fig:c0merged}
\end{figure}

Besides the transfer velocity, also the boundary value $c_0$ of the concentration is an important control parameter influencing the pattern formation process. The dependence of the wavenumber $k$ in the patterning regime on the transfer velocity $v$ and the boundary concentration $c_0$ is shown in Figure \ref{fig:c0merged}. One can clearly see that the patterning regime broadens and shifts towards higher velocities $v$ for increasing $c_0$. Without loss of generality, $c_0=-0.9$ will be used for the following discussions.

\section{Synchronization with periodic prestructures in 1D}
\begin{figure}[ht]
  \centering
 \includegraphics[width=0.47\textwidth]{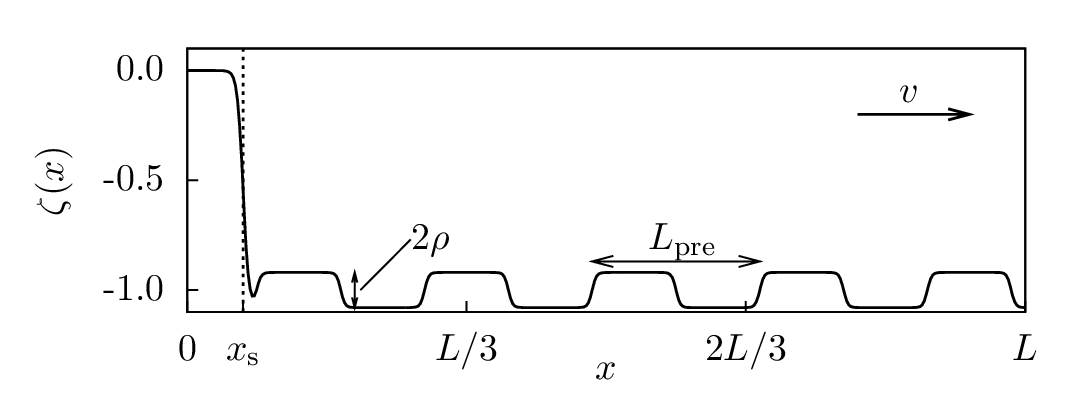}
 \caption{Sketch of the spatial dependence of $\zeta(x)$ defined by Eqn.\ \eqref{eq:zeta}-\eqref{prestructure} for a prestructured substrate, indicating the local strength of the SMC. The sketch is only a snapshot as the prestructure moves with the velocity $v$ to the right.}
 \label{fig:zetarho}
\end{figure}

In general, two different types of prestructures can be distinguished. Topological prestructures consist of the same material as the underlying substrate and are defined by the spatially varying height-profile. In contrast, chemically prestructured substrates have a flat profile but another property that is spatially varying, like, e.g., the wettability. In experiments, both types often occur simultaneously, when prestructures made of a different material than the substrate are used, e.g. gold stripes on a silicon substrate. As we assume no dynamics of the liquid layer in our model (\ref{eq:model})-(\ref{eq:boundary_conditionsx}), only the varying interaction of the monolayer with the substrate has to be included, which is connected to the disjoining pressure \cite{thiele2006driven,thiele2003modelling}. Therefore one can model the prestructure via a spatial modulation $m(\mathbf{x},t)$ of the strength $\zeta(\mathbf{x})$ of the SMC,
\begin{equation}
 \zeta(\mathbf{x}) = -\frac{1}{2}\left( 1+\tanh \left( \frac{x-x_\mathrm{s}}{l_\mathrm{s}}\right) \right)\left( 1+\rho \, m(\mathbf{x},t)\right).
\label{eq:zeta}
\end{equation}
The form of the function $m(\mathbf{x},t)$ mimics the form of the prestructure. For stripes that are parallel to the meniscus we use
\begin{equation}
m(\mathbf{x},t)=  \tanh  \left( 10 \left( 4 \left| \mathrm{frac} \left( \frac{x-vt}{L_\mathrm{pre}}  \right)-0.5 \right|-1 \right) \right).
\label{prestructure}
\end{equation}
This delivers a kink-antikink train with periodicity $L_\mathrm{pre}$, where the steepness of the kinks is determined by the constant $a$ and $\mathrm{frac}$ denotes the fractional part of the argument. The strength of the prestructure is determined by the contrast $\rho$. A sketch of a resulting strength $\zeta(x)$ of the SMC in one dimension is shown in Figure \ref{fig:zetarho}. Note that the prestructure is fixed to the substrate and therefore also moves with a velocity $v$ in the reference frame. \par 
The resulting model \eqref{eq:model}-\eqref{prestructure} is a spatially forced Cahn-Hilliard equation, which has large similarities to studies from the literature \cite{Krekhov2004Phase,Krekhov2009Formation,weith2009traveling}. In \cite{Krekhov2004Phase}, a uniformly quenched Cahn-Hilliard model influenced by a resting spatial forcing is studied, while in \cite{Krekhov2009Formation} a Cahn-Hilliard model subject to a moving quenching front was discussed. In the latter, also a modulation of the quenching front position is included. An extension of \cite{Krekhov2004Phase} towards a moving spatial forcing is presented in \cite{weith2009traveling}. The model \eqref{eq:model}-\eqref{prestructure} incorporates aspects of all the work mentioned, as the pattern formation occurs at a fixed front defined by the onset of SMC at $x=x_\mathrm{s}$ and is influenced by a moving spatial forcing, while the whole system is also subject to an advection. Although similar, the front in the model \eqref{eq:model}-\eqref{prestructure} is not equal to the quenching front present in \cite{Krekhov2009Formation,weith2009traveling}. A quenching front typically describes the transition from a one phase (single-well potential) to a two phase (double-well potential) region, while the front used here is a transition from a symmetric double-well potential to a tilted double-well potential. Additionally, the fixed front in combination with the advection is a crucial aspect of this model, as it selects a distinct wavelength of the emerging pattern, in contrast to a classical Cahn-Hilliard model, where no distinguished wavelength exists. That is, the pattern formation process in the case described here is driven by the concentration oscillations at the meniscus, which are subject to the moving prestructure, creating a temporal periodic forcing. Therefore, one can think of the occurring patterns as the result of a synchronization process.\par
\begin{figure}[ht]
  \centering
 \includegraphics[width=0.5\textwidth]{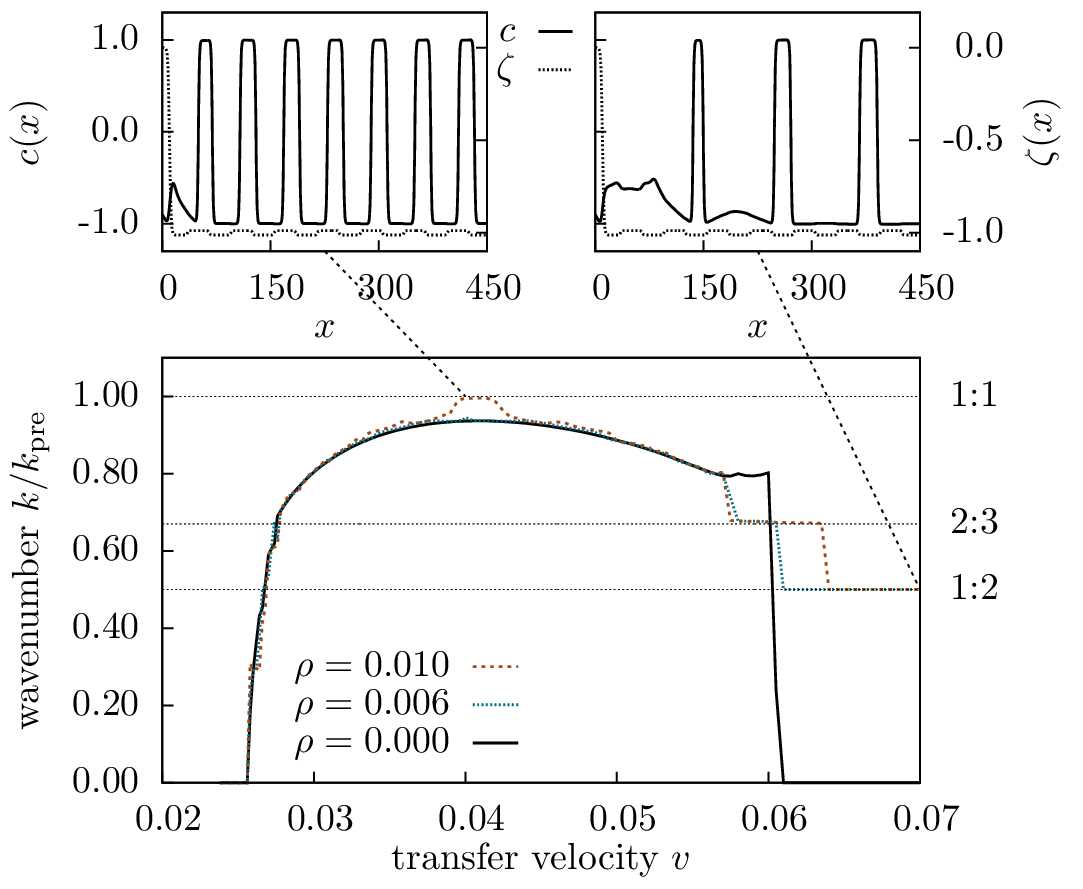}
  \caption{Wavenumber $k$ of generated patterns during a transfer onto a prestructured substrate in one dimension in dependence on the transfer velocity $v$ and the prestructure contrast $\rho$ of a prestructured substrate with $L_\mathrm{pre}=60$ (bottom). Two snapshots of exemplary solutions for $v=0.04$ and $v=0.07$ (both at $\rho=0.01$) are shown (solid lines, top), indicating 1:1 and 1:2 synchronization with the prestructure (dotted lines, top).}
 \label{fig:k-vs-v-pre}
\end{figure}

In contrast to the work presented in \cite{koepf2012substrate}, we now investigate the transfer onto a prestructured substrate and therefore a nonvanishing contrast $\rho \neq 0$. In this case the $k$ versus $v$ curve changes by exhibiting jumps to plateaus that correspond to a wavenumber commensurable with the wavenumber of the prestructure $k_\mathrm{pre}$. These plateaus grow with increasing contrast $\rho$, as can be seen in Figure \ref{fig:k-vs-v-pre}. Within these plateaus, the generated pattern is almost independent of the transfer velocity, which can reduce the necessary accuracy of the transfer velocity control in the experimental system. In addition, the use of a prestructured substrate extends the velocity range in which periodic structures are generated towards higher transfer velocities, as well as extends the wavenumber range that is accessible towards higher wavenumbers, i.e. smaller wavelengths, which can be seen at the 1:1 synchronization plateau. Both effects enable the production of a 
broader range of patterns at a larger range of experimental parameters. Note that the wavenumbers $k$ plotted in Figure \ref{fig:k-vs-v-pre} are averaged over time, because for higher order synchronization ratios, like 2:3, the resulting pattern can consist of alternating stripes with different periodicities, which only in average are commensurable with the prestructure periodicity. \par
\begin{figure}[ht]
  \centering
 \includegraphics[width=0.5\textwidth]{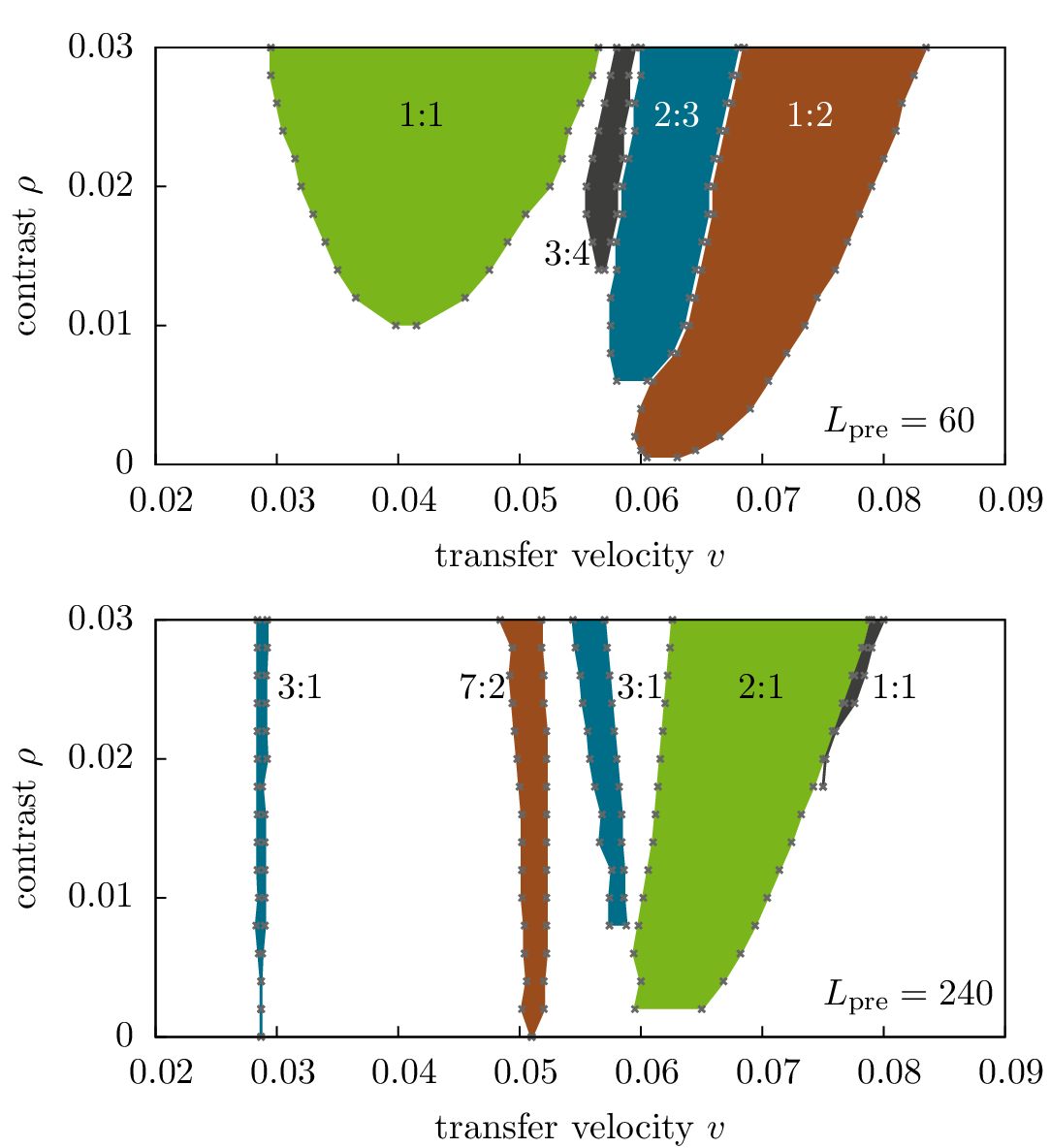}
 \caption{Arnold tongue diagrams showing the synchronization regimes depending on transfer velocity and prestructure contrast, where different colors correspond to different synchronization ratios. The crosses depict the boundaries of the synchronization regimes. The wavelength of the prestructure is $L_\mathrm{pre}=60$ (top) and $L_\mathrm{pre}=240$ (bottom), corresponding to a wavenumber $k_\mathrm{pre}=\frac{2\pi}{L_\mathrm{pre}}\approx 0.105$ (top) and $k_\mathrm{pre}\approx 0.026$ (bottom).}
 \label{fig:Arnold}
\end{figure}

A good overview of the possible synchronization regimes can be gained in an Arnold tongue diagram \cite{pikovsky2003synchronization}, where the parameter regions in which synchronization occurs are marked as colored areas in the $\rho$-$v$ plane, with each color referring to a different synchronization ratio (see Figure \ref{fig:Arnold}). Of course the structure of such a diagram also depends on the periodicity of the prestructure as can be seen in the comparison of the Arnold diagrams for a prestructure wavelength of $L_\mathrm{pre}=60$ (top) and $L_\mathrm{pre}=240$ (bottom). However, common features can be identified, like the shape of the 1:2 synchronization tongue in the $L_\mathrm{pre}=60$ diagram and the 2:1 synchronization tongue in the $L_\mathrm{pre}=240$ diagram, which both correspond to stripes with a periodicity of $\lambda=120$.\par
Despite obvious quantitative differences, the results obtained in the model (\ref{eq:model})-(\ref{prestructure}) are similar to the results obtained in the full two-component model for Langmuir-Blodgett transfer presented in \cite{koepf2011controlled}. Both models predict the existence of similarly shaped synchronization regimes for various synchronization ratios at a broad range of transfer velocities and prestructure contrasts. The feature of increasing synchronization domains for increasing prestructure contrast $\rho$ is also common to both models.

\section{Effects of periodic prestructures in 2D}
\begin{figure}[ht]
  \centering
 \includegraphics[width=0.48\textwidth]{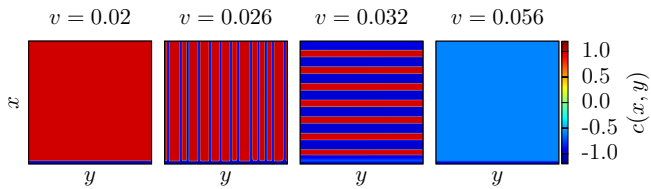}
 \caption{Overview of basic pattern types in two dimensions on homogeneous substrates ($\rho=0, L =600$). The meniscus is located at the bottom of the snapshots, with the transfer direction going from bottom to top. The patterns correspond to a homogeneous LC transfer, stripes perpendicular to the meniscus, stripes parallel to the meniscus and a homogeneous LE transfer (from left to right).}
 \label{fig:2Dhomogeneous}
\end{figure}

The stable solution types of the one-dimensional system, i.e. homogeneous transfer of LE layers, alternating stripes in the LE and LC phase parallel to the meniscus, and homogeneous transfer of LC layers, are also solutions of the two-dimensional system (\ref{eq:model})-(\ref{eq:boundary_conditionsy}), if they are only extended homogeneously in the new spatial $y$ direction \cite{koepf2012substrate}. Furthermore, for low transfer velocities within the patterning regime, stripes parallel to the meniscus are unstable and stripes perpendicular to the meniscus are created and transferred, see Figure \ref{fig:2Dhomogeneous} for an overview of the possible two-dimensional basic pattern types. \par
\begin{figure}[ht]
  \centering
 \includegraphics[width=0.48\textwidth]{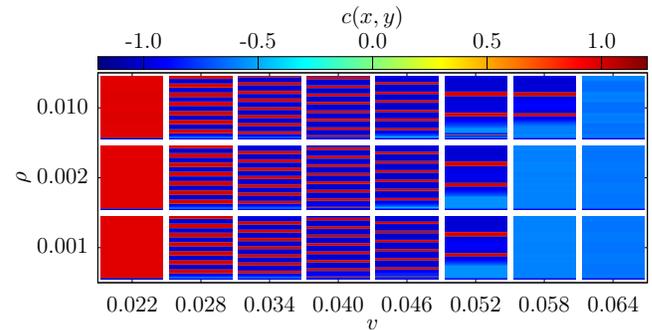}
 \caption{Overview of occurring patterns on a prestructured substrate with stripes parallel to the meniscus with $L_\mathrm{pre} = 200$.}
 \label{fig:2DSyncOverviewx}
\end{figure}

Considering prestructured substrates in two dimensions, the results from the one-dimensional system can be directly reproduced by using stripes parallel to the meniscus as a prestructure. Snapshots of numerical solutions for different transfer velocities $v$ and prestructure contrasts $\rho$ are shown in Figure \ref{fig:2DSyncOverviewx}, where only solutions homogeneous in $y$ direction occur. This means that stripes parallel to the meniscus are further stabilized by a prestructure with stripes parallel to the meniscus. This is particularly important in the lower velocity part of the patterning regime, where a transfer onto a homogeneous substrate would result in stripes perpendicular to the meniscus. That is, the instability leading to this solution type is suppressed by the use of a prestructured substrate. Interestingly, there also exist small parameter regimes (e.g. $v=0.038,\ \rho=0.01$, not shown), where the pseudo-1D patterns are not further stabilized by a pseudo-1D prestructure, but in fact 
destabilized, so that the stripes parallel to the meniscus break up into smaller domains, which are still roughly aligned in lines. This is reminiscent of similar results found in the Swift-Hohenberg equation \cite{manor2008wave}, where a pseudo-1D forcing can also destabilized equally aligned stripe patterns, leading to oblique or rectangular patterns. \par
\begin{figure}[ht]
  \centering
 \includegraphics[width=0.48\textwidth]{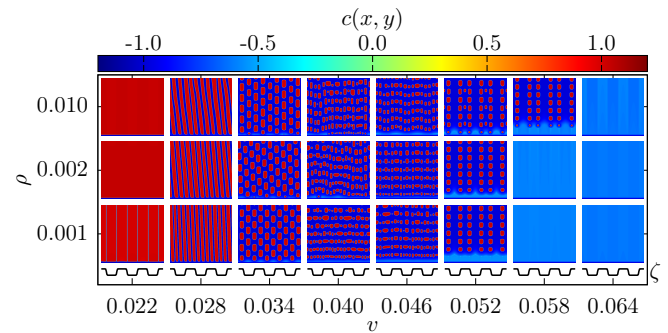}
 \caption{Overview of occurring patterns on a prestructured substrate with stripes perpendicular to the meniscus with $L_\mathrm{pre} = 200$. A schematic cross section of the prestructure is shown below the solution panels. The possible patterns comprise stripes with different orientations as well as regular and irregular lattice structures.}
 \label{fig:2DSyncOverviewy}
\end{figure}
In two dimensions the prestructure can also be oriented differently, e.g. perpendicular to the meniscus. We consider a prestructure of the form 

\begin{equation}
 m(\mathbf{x})=  \tanh  \left( 10 \left( 4 \left| \mathrm{frac} \left( \frac{y}{L_\mathrm{pre}}  \right)-0.5 \right|-1 \right) \right).
\end{equation}

\begin{figure}[ht]
  \centering
 \includegraphics[width=0.48\textwidth]{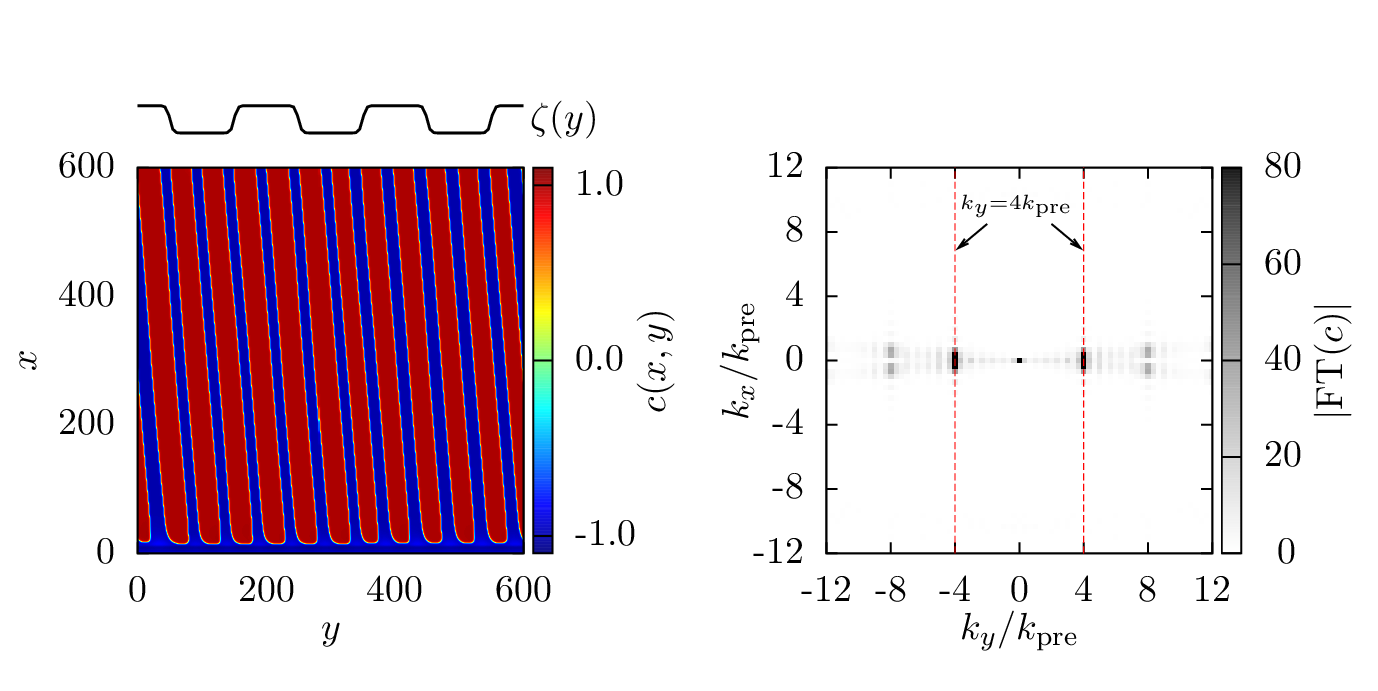}
 \caption{Oblique stripe pattern arising on a vertically prestructured substrate ($L_\mathrm{pre} = 200,\ v= 0.028,\ \rho=0.002$) and its Fourier transform. The dominant Fourier modes are integer multiples of the wavenumber $k_\mathrm{pre}$ of the prestructure.}
 \label{fig:Fouriertilted}
\end{figure}

The transferred patterns are shown in Figure \ref{fig:2DSyncOverviewy}. In this case a large variety of qualitative different structures can be generated depending on the transfer velocity $v$ and prestructure contrast $\rho$. For low velocities ($v=0.022$, $\rho=0.001$), stripes perpendicular to the meniscus are formed, just like in the case of a homogeneous substrate, but in a synchronized manner with wavelengths that have a fixed 2:1 ratio to the wavelength of the prestructure ($L_\mathrm{pre}=200$). For increased velocities ($v=0.028$), stripes that are slightly tilted against the prestructure are created. There, the prestructure wavelength or a commensurable ratio of it are no favorable wavelengths for system. Therefore the system effectively changes the wavelength by tilting the stripes -- similar to the Benjamin-Feir instability mechanism leading to the zigzag pattern in other pattern forming systems \cite{cross1993pattern} -- while reacting to the prestructure with the $y$-component of the wave 
vector. In the case of $v=0.028,\ \rho=0.002$, the $y$-component exhibits a 4:1 synchronization, which can be best seen looking at the Fourier transform of the pattern (see Figure \ref{fig:Fouriertilted}, right panel). While the tilt angle is defined by the ratio of the wavelength the system favors and the wavelength introduced by the prestructure, the tilt direction (to the left or the right) is determined by the initial conditions, and is sensitive to slight perturbations. \par
Further increased velocities lead to more complex pattern topologies. For $v=0.034$, lattice structures of small domains in the LC phase are created, where each two consequent rows are shifted horizontally by half a wavelength. For higher velocities, this does not hold true, and the LC domains of the patterns are not regular any more, resulting in irregular looking patterns. However, for some parameter sets, quite regular rows consisting of irregular domains can be identified, e.g., for $v=0.046,\ \rho=0.002$. For high velocities near the upper boundary of the patterning regime, fully regular patterns arise again. They consist of domains that are well aligned in rows and columns and are synchronized to the prestructure. They can be understood as a superposition of the natural pattern in the absence of a prestructure, which are stripes parallel to the meniscus, and the perpendicular stripe pattern induced by the prestructure.\par
The comparison of Figures \ref{fig:2DSyncOverviewx} and \ref{fig:2DSyncOverviewy} exhibits completely different behavior of the patterning process, only depending on the orientation of the prestructure. This reveals the different nature of the two basic pattern types on homogeneous substrates, which are stripes parallel and perpendicular to the meniscus. Stripes perpendicular to the meniscus occur after a secondary instability, after stripes parallel to the meniscus have been formed \cite{koepf2012substrate}. Therefore this instability might be suppressed more easily by an according prestructure (see Figure \ref{fig:2DSyncOverviewx}) than the instability leading to stripes parallel to the meniscus. This could explain the various patterns shown in Figure \ref{fig:2DSyncOverviewy}, which might result from a competition between the strong tendency to form stripes parallel to the meniscus for high transfer velocities and the prestructure that favors stripes perpendicular to the meniscus.

\section{Conclusions and outlook}
A theoretical investigation of Langmuir-Blodgett transfer onto prestructured substrates by means of a generalized Cahn-Hilliard model has been presented. Starting with the one-dimensional case, we found synchronization effects of different order with periodic prestructures. Utilizing these effects, the patterning process can be controlled more precisely, new patterns of higher complexity can be obtained, and the patterning regime can be extended to a larger control parameter range. \par
In the two-dimensional case, prestructured substrates can be used to stabilize the production of stripes parallel to meniscus, as well as to enable a variety of different complex patterns, if the prestructure orientation is changed. Again, synchronization effects enable an additional control mechanism over the patterning process. This concept might be extended to similar systems, like orientation control in a quenched system \cite{weith2009traveling}.\par
Towards a more detailed understanding of the processes in the real experiments one should further investigate the legitimacy of the assumptions made in the derivation of the Cahn-Hilliard model used here, in the case of prestructured substrates. The approximation of a static meniscus might become improper for prestructures, which introduce a wettability contrast, and therefore influence the dynamics of the meniscus. This falls out of the scope of this paper and will be the topic of future work.


%
%

%

\begin{acknowledgments}
This work was supported by the Deutsche Forschungsgemeinschaft within SRF TRR 61. We thank Michael H. K\"opf, Lifeng Chi and Uwe Thiele for fruitful discussions. We also gratefully acknowledge many fruitful discussions with the late Prof. Dr. Rudolf Friedrich.
\end{acknowledgments}

\bibliographystyle{unsrt}
\bibliography{Literature}

\end{document}